\documentclass[fleqn,usenatbib]{mnras}

\usepackage{newtxtext,newtxmath}
\usepackage[T1]{fontenc}

\DeclareRobustCommand{\VAN}[3]{#2}
\let\VANthebibliography\thebibliography
\def\thebibliography{\DeclareRobustCommand{\VAN}[3]{##3}\VANthebibliography}

\usepackage{graphicx}	% Including figure files
\usepackage{amsmath}	% Advanced maths commands
\usepackage{bm}	% Advanced maths commands
\title[X-ray polarisation from disc winds]{X-ray polarisation properties of thermal-radiative disc winds in binary systems}

\author[R. Tomaru, C. Done, H. Odaka]{
Ryota Tomaru$^{1}$\thanks{E-mail: ryota.tomaru@durham.ac.uk}
Chris Done$^{1,2}$
Hirokazu Odaka$^{3}$
\\
$^{1}$Centre for Extragalactic Astronomy, Department of Physics, Durham University, South Road, Durham DH1 3LE, UK\\
$^{2}$Kavli Institute for Physics and Mathematics of the Universe (WPI), University of Tokyo, Kashiwa, Chiba 277-8583, Japan\\
$^{3}$Department of Earth and Space Science, Graduate School of Science, Osaka University, 1-1 Machikaneyama, Toyonaka, Osaka 560-0043, Japan
}

\date{Accepted XXX. Received YYY; in original form ZZZ}

\pubyear{2023}

\begin{document}
\label{firstpage}
\pagerange{\pageref{firstpage}--\pageref{lastpage}}
\maketitle

% Abstract of the paper
\begin{abstract}

New X-ray polarisation results are challenging our understanding of the accretion flow geometry in black hole binary systems. 
Even spectra dominated by a standard disc can give unexpected results, such as the high inclination black hole binary 4U 1630-472, where the observed X-ray polarisation is much higher than predicted. 
This system also shows a strong, highly ionised wind, consistent with thermal-radiative
driving from the outer disc, leading to speculation that scattering in the wind is responsible for the unexpectedly high polarisation degree from a standard optically thick disc. 
Here we show that this is not the case.
The optically thin(ish) wind polarises the scattered light in a direction orthogonal to that predicted from a standard optically thick disc, reducing about 2\% rather than enhancing the predicted polarisation of the total emission. 
This value is consistent with the polarisation difference between the disc-dominated soft state, where absorption lines by the wind are clearly seen, and the steep power-law state, where no absorption lines are seen. 
If this difference is genuinely due to the presence or absence of wind, the total polarisation direction must be orthogonal to the disc plane rather than parallel as expected from optically thick material. 

\end{abstract}

% Select between one and six entries from the list of approved keywords.
% Don't make up new ones.

%%%%%%%%%%%%%%%%%%%%%%%%%%%%%%%%%%%%%%%%%%%%%%%%%%

%%%%%%%%%%%%%%%%% BODY OF PAPER %%%%%%%%%%%%%%%%%%

\section{Introduction}\label{sec:intro}
% Example figure

The NASA-ASI Imaging X-ray Polarimetry Explorer ({\it IXPE}, \citealt{Weisskopf2016}) has opened a new window for exploring the geometry of X-ray emission regions by detecting the net direction of the electric vector of the X-ray emission. A non-zero polarisation implies a non-spherical 
geometry, e.g., an accretion disc.
The classic result for a plane-parallel, optically thick, electron-scattering atmosphere (approximating a standard disc) 
gives a polarisation degree (PD) of zero when viewed directly from above (as the source is symmetric when face-on) to $11.7$\% for completely edge-on \citep{Chandrasekhar1960}.

Emission from a disc around a black hole is affected by strong gravity. 
Geodesic transfer of the light rays rotates the plane of polarisation, depolarising the radiation from the innermost regions, which emit at the highest temperatures
\citep{Connors1980, Dovciak2008, Mikusincova2023},
giving a diagnostic of black hole spin. 
This, plus the non-zero scale height of the outer disc, which blocks the highest inclination angles, means that the maximum PD expected from an accretion disc is of order 5-6\%. 

{\it IXPE} observed the black hole binary (BHB) 4U1630-472 (hereafter 4U1630) in its recent outburst in 2022-23 and detected strong polarisation.
The system parameters of this source are not well known as it is at 
a large distance and is strongly absorbed \citep{Kalemci2018}. Nonetheless, the inclination can be estimated as  $65^{\circ} < i < 75^{\circ}$ from the combination of the presence of dips, and the absence of eclipses \citep{Kuulkers1998, Tomsick1998}.
The high inclination is also implied by the detection of highly ionised absorption lines in its spectrum  \citep{Kubota2007, DiazTrigo2014, Neilsen2014, Hori2018, Gatuzz2019} which arise from the line of sight intercepting an equatorial disc wind \citep{ponti2012}.
Thus, we expect to see polarisation from an accretion disc with PD of $\approx 5$\%, potentially decreasing at the highest energies due to the general relativistic effects. 
Instead, the data show polarisation, increasing with energy from 6\% at 2~keV to 10\% at 8~keV \citep[hereafter R23]{Rawat2023, Kushwaha2023, Ratheesh2023}. 
This motivated speculation that the wind contributes to the observed polarisation \citep{Rawat2023, Kushwaha2023}.
The wind should give a polarisation signal as it has a preferentially equatorial geometry from its disc origin and scatters some fraction of the incident flux.

In our previous work, we made a physical model of these winds from irradiation of the outer disc by the X-ray hot inner disc. The heated material expands and forms a wind driven by the pressure gradient for radii where the sound speed exceeds the escape velocity (thermal winds: \citealt{Begelman1983a, Done2018}).
We used a radiation-hydrodynamic code to calculate the geometry and kinematics of this material, also including the 
additional push by radiation pressure (thermal-radiative winds: \citealt{Tomaru2019}). 
We used the density/velocity structure as input to a  
Monte Carlo radiation transfer code, {\sc monaco} v1.6.0 \citep{Odaka2011}, to predict detailed line profiles, and showed that these matched well to those observed
from both neutron star \citep{Tomaru2020b,Tomaru2023b} and black hole binaries \citep{Tomaru2020}.
This includes the anomalous wind seen in the BHB GRO J1655-40, which was previously claimed to require magnetic driving but where the launch radius was underestimated due to the impact of radiative cascades and optical depth effects on the density diagnostic line  \citep{Tomaru2023a}.

Here we extend {\sc monaco} to handle polarisation in radiation transfer and use this to explore the polarisation properties of a thermal-radiative wind.  
We show that the wind
{\it reduces} the total polarisation, 
as scattering from the optically thin equatorial wind polarises in the opposite direction (perpendicular to the disc plane, i.e. parallel to the spin axis
) than that expected for the intrinsic 
disc emission (parallel to the disc plane: see, e.g. \citealt{Sunyaev1985}, hereafter ST85), so the wind worsens the mismatch.

\section{Calculating polarisation in {\sc Monaco}}\label{sec:monaco}

\begin{figure}
    \centering
    \includegraphics[width=0.9\hsize]{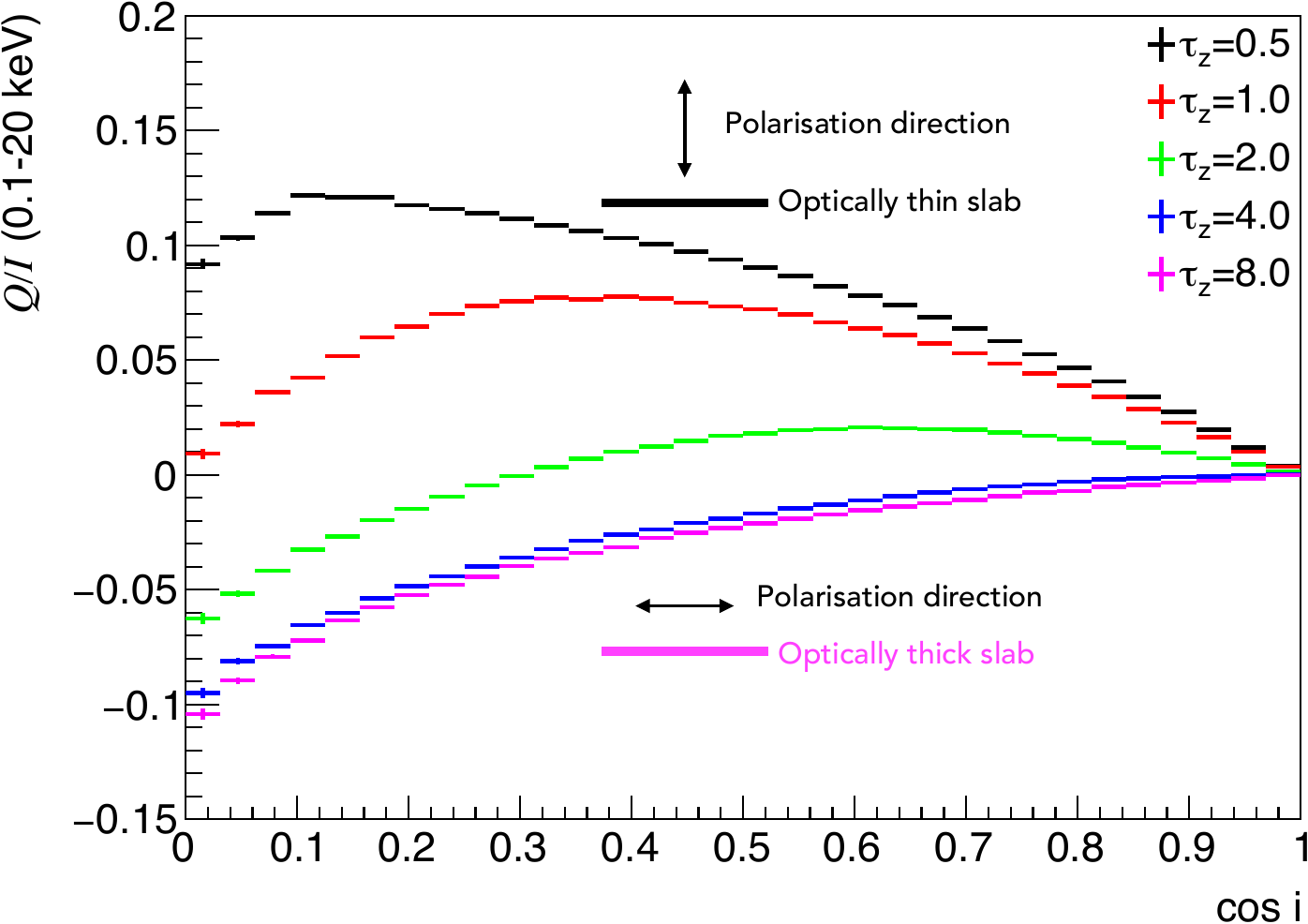}
    \caption{The polarisation degree produced by electron scattering from a slab of different optical depths. This matches the results of ST85, though we caution that they use the opposite sign convention. We have included sketch geometries to make it clear that the polarisation expected from a standard optically thick disc is aligned along the disc plane (defined as negative). In contrast, the polarisation expected from an optically thin slab is perpendicular to the disc, i.e. aligned with the radio jet (defined as positive). }
    \label{fig:slab}
\end{figure}

We use a Monte Carlo simulation code, {\sc monaco}  \citep{Odaka2011}, for this study.
This is a general-purpose framework for synthesising X-ray radiation from astrophysical objects with complex geometry by calculating radiative transfer based on a Monte Carlo approach.
It includes a full treatment of Comptonisation from electrons with non-relativistic thermal and bulk motions, which is fully tested as 
described in \citet{Odaka2014}.

We incorporate a new module to track photon polarisation.
We adopt the algorithm given by \citet{Matt1996} to calculate the 
polarisation vector (a unit vector parallel to the electric field) of a Compton scattered photon. 
We consider the scattering process of an initial photon in the electron's rest frame of reference, 
so the observer's frame is related to this frame through the Lorentz transformations.
The photon 
energy $E$, direction $\hat{\bm{k}}$, and polarisation vector $\hat{\bm{e}}$ before (0) and after (1) scattering
can be written as
\begin{equation}
(E_0,\ \hat{\bm{k}}_0,\ \hat{\bm{e}}_0)\to(E_1,\ \hat{\bm{k}}_1,\ \hat{\bm{e}}_1),
\end{equation}
The degree of the polarisation of the scattered photon is given by
\begin{equation}
p = \frac{2-2\sin^2\theta\cos^2\phi}{\dfrac{E_1}{E_0}+\dfrac{E_0}{E_1}-2\sin^2\theta \cos^2\phi},
\end{equation}
where $\theta$ and $\phi$ denote the scattering and azimuth angles measured from the initial polarisation vector $\hat{\bm{e}}_0$, respectively.
The polarisation vector $\hat{\bm{e}}_1$ of the scattered photon is assigned to be
\begin{gather}
\hat{\bm{e}}_1 = \frac{1}{A_1}\hat{\bm{k}}_1\times (\hat{\bm{e}}_0\times\hat{\bm{k}}_1) = \frac{1}{A_1}(\hat{\bm{e}}_0-(\hat{\bm{e}}_0\cdot{\hat{\bm{k}}_1})\hat{\bm{k}}_1), \\
A_1 = |\hat{\bm{k}}_1\times (\hat{\bm{e}}_0\times\hat{\bm{k}}_1)|
\end{gather}
with a probability equal to the polarisation fraction $p$; $A_1$ is merely the size of the calculated vector for normalisation.
Otherwise, the photon is depolarised with a probability of $1-p$; therefore, the polarisation vector is assigned to be randomly sampled.
This treatment results in unpolarised photons after a number of Monte Carlo trials.

We first demonstrate the performance of the polarisation module in {\sc monaco} by showing results for electron scattering in a constant density plane parallel slab with optical depth from $\tau=0.5-8$ (see Fig.\ref{fig:slab}) to compare with the classic paper of ST85.
ST85 assumed Thompson scattering (no electron energy change) of seed photons with energy $\ll kT_e$, but {\sc monaco} always includes the full Compton scattering cross-section. 
We set the seed photons as a blackbody distribution at
the radiation temperature ($kT_{\rm e}=kT_{\rm bb}=1~{\rm keV}$) so that up-scattering and down-scattering cancel after multiple events. 
We emit $4\times 10^8$
seed photons isotropically on the midplane of the slab to compare to geometry (a) in ST85.
We use $-1< \cos i< 1$ for all computations but only show the $0<\cos i<1$ in all figures due to the symmetry in this paper.

Fig.\ref{fig:slab} shows the resulting polarisation (Stokes parameters $Q/I$ as
$U=0$ due to axisymmetry) as a function of inclination. These are 
integrated over the entire spectrum 
to compare directly with ST85 (their Fig 6a). 
However, we use the opposite sign convention here, so that the positive $Q$ direction is normal to the disc (or slab) plane, as shown by the inset sketch geometries. This sign convention is the same as \citet{Chandrasekhar1960}. Our results match well with ST85, 
demonstrating that the polarisation is calculated correctly, with the highest optical depths tending towards the classical \citet{Chandrasekhar1960} result, while lower optical depths have lower polarisation, which changes sign 
as the scattering medium goes optically thin. 

Physically, the change in the sign of polarisation is due to the corresponding change in the direction of the photons before their final scattering.
In an optically thick slab, the photons are all diffusing upwards, so they have fairly uniform momentum vectors perpendicular to the plane. 
Hence, they have an electric vector which is parallel to the plane.
Electron scattering tends to preserve the plane of the electric vector,
so the photon resulting from the final scattering also has an electric vector parallel to the plane of the slab (defined as a negative sign on Q).
By contrast, in the optically thin slab, photons scattered by an electron are more likely to be moving parallel to the plane of the slab as they are the ones which encounter the largest optical depth and have the largest probability of being scattered. These can have a range of azimuthal angles with respect to the observer.
Those which are at an azimuth of 0 (directly towards) or 180 (directly away) will have electric vectors before (and after) scattering in the plane of the slab,
but the majority are at other azimuths, which 
gives a net electric vector perpendicular to the slab plane (positive polarisation) \citep{Angel1969}.

This switch in polarisation direction with optical depth is important to understand what is expected from BHB as they show a state transition,
from being dominated by an optically thick,
geometrically thin disc (high/soft state), to the low/hard state where the emission is instead dominated by Comptonisation by hot  ($\sim 100$~keV), optically thin plasma.
If this hot accretion flow has  a disc-like geometry,
then this predicts that the state transition accompanies a change in the sign of the polarisation. 
%the state change is accompanied by a change in sign of the polarisation,
with the high/soft state polarisation aligned with the plane of the disc, which is perpendicular to the jet,
to the low/hard state aligned perpendicular to the plane of the disc,
which is parallel to the jet.
We note that the net polarisation predicted for the hot flow will be smaller than that calculated in Fig.\ref{fig:slab} due to the larger energy change in inverse Compton scattering in the hot ($\sim 100$~keV) flow which reduces the polarisation (Eq. 2) compared to that calculated here for electrons at 1~keV. 

{\it IXPE} has observed a low/hard state in the BHB Cyg X-1 \citep{Krawczynski2022}. 
To zeroth order, this follows the expectations above for an optically thin(ish) flow in a disc geometry in that the polarisation is aligned with the jet (i.e. perpendicular to the disc). 
This rules out multiple other geometries where the optically thin material is a compact source on the spin axis (almost no polarisation) or aligned along the jet direction (polarisation with the wrong sign),
though a more detailed understanding of the polarisation and hence source geometry is still missing\citep{Krawczynski2022, Veledina2023, Zdziarski2023}.
We note that very similar polarisation, aligned with the jet, is seen in the low Eddington fraction AGN NGC4151 \citep{Gianolli2023}, where the accretion flow is likely to be similar to the low/hard state in black hole binaries \citep{Kubota2018, Mahmoud2020}.

By contrast, the disc-dominated state is expected to have polarisation parallel to the disc plane, i.e., perpendicular to the jet direction. 
However, this expected switch has not been seen in the soft state of Cyg X-1; 
instead, the polarisation direction is consistent with that of the hard state, which is parallel to the jet axis, and its magnitude is $\sim 2\%$ \citep{Dovciak2023}. It is not clear whether this is typical for disc dominated states in general. Cyg X-1 may be a special case as it never makes a very clean transition to the disc dominated state \citep{Done2005, Sugimoto2016}. 
This could be due to it never quite reaching high enough luminosity to make a full transition or that there is always some component of wind-fed accretion from the high mass companion star which prevents the disc from forming. 

These results already show that optically thin scattering by a wind is likely to depolarise the total emission from an optically thick accretion disc as they give opposite polarisation directions. We explore this further below, using a physical model for the wind in the high/soft state of 4U1630.

\section{Polarisation from a wind} \label{sec:pol}
\subsection{The wind model}

\begin{figure}
    \centering
    \includegraphics[width=0.9\hsize]{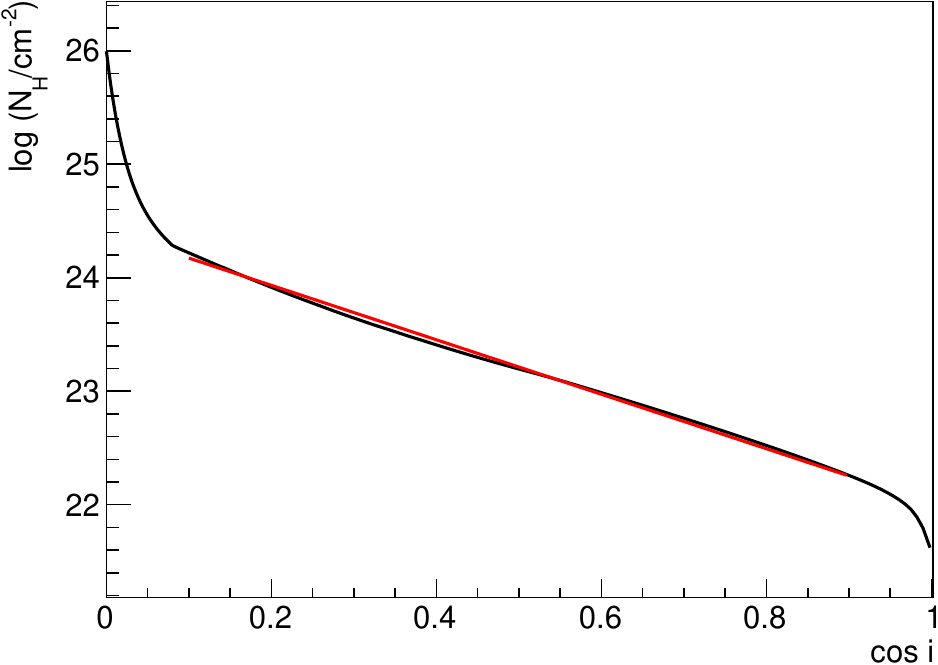}
    \caption{The comparison between column density of the accretion disc wind from the RHD simulation and the analytic fitting formula. The disc equatorial plane is at $\cos i=0$. }
    \label{fig:column}
\end{figure}
\begin{figure}
    \centering
    \includegraphics[width=0.9\hsize]{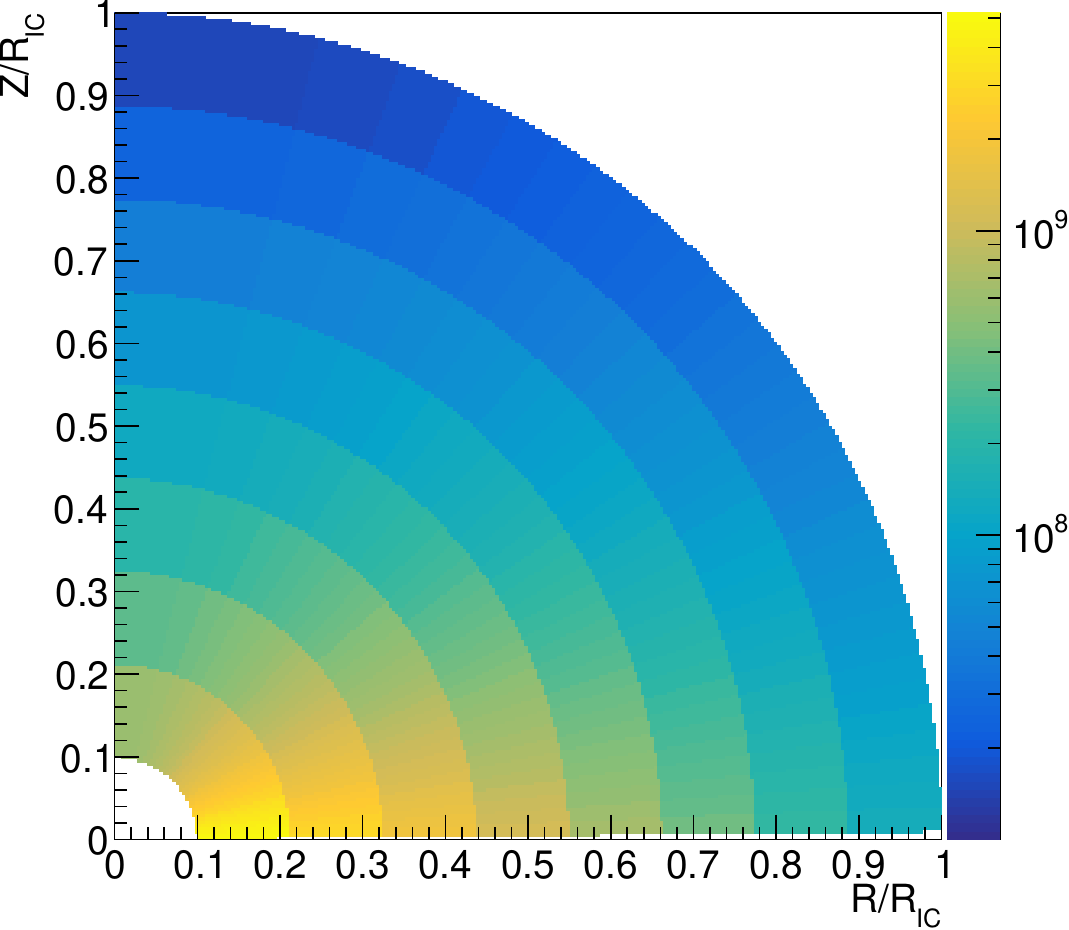}
    \caption{The density distribution calculated by Eq.\ref{eq:density}.}
    \label{fig:density}
\end{figure}
Thermal-radiative winds depend on the spectral shape (setting the radiation temperature), total luminosity (setting the heating rate and radiation pressure) and disc size (e.g. \citealt{Done2018}).
Only the first of these is securely known from the data for 4U1630, as the high optical extinction prevents observations of the companion star, so neither distance nor orbit is known. 

Instead, we use our previous simulation from the similarly highly inclined neutron star binary GX13+1 \citet{Tomaru2020b}. This has similarly large equivalent widths in highly ionised iron absorption lines to those seen in 4U1630 ($\sim 50$~eV for Fe {\sc xxvi}, and $25-30$~eV for Fe {\sc xxv} \citealt{Gatuzz2019} and \citealt{Allen2018}), so we use this wind as the basis of our polarisation calculation. 
This simulation has Compton temperature $T_{\rm IC}= 1.3\times 10^7$~K, giving a Compton radius 
(where the typical thermal speed of the particles exceeds the local escape velocity) 
of $R_{\rm IC}=0.61 m_p G M /(k T_{\rm IC}) 
\sim (6\times 10^5/T_{\rm IC,7}) R_g$ 
where $T_{\rm IC,7}=T_{\rm IC}/10^7~K$.
Other important parameters are luminosity $L=0.5L_{\rm Edd}$ and outer disc radius $10R_{\rm IC}$.

Fig. \ref{fig:column} shows the resulting column density of the radiation-hydrodynamic simulation as a function of viewing angle (black: from Fig 2 in  \citealt{Tomaru2020b}). This can be well described by an analytic fitting function $\log (N_{H}(\mu)/{\rm cm^{-2}})=22+2.4(1-|\mu|)$ in the range of $0.1 < |\mu| < 0.9$, where $\mu=\cos i$ (Fig. \ref{fig:column}: red line). 
This is a typical column density distribution in the thermal wind models, giving a much better description of the radiation-hydrodynamic results than the $N_H\propto (1-|\mu|)$ assumed in \citet{Done2018}. 
The system parameters of 4U1630 are probably rather similar in terms of Compton temperature, luminosity and $R_{\rm out}/R_{\rm IC}$, so the wind should also be similar.

We make an analytic approximation to this density distribution, $n(R, \mu)$, such that
\begin{equation}
\label{eq:density}
n(R, \mu) = N_{\rm H}(\mu)\frac{R_{\rm in}R_{\rm out}}{(R_{\rm out}-R_{\rm in})R^2}
\end{equation}
where the $n\propto R^{-2}$ dependence comes from the launch condition for thermal winds \citep{Done2018}. 

We sample this density structure on a logarithmic radial grid ($\Delta R/R$ constant: 8 bins from $R_{\rm in}-R_{\rm out}$ of $0.1-1 R_{\rm IC}$, where the outer disc radius is smaller than the original radiation hydrodynamic simulation to reduce computational cost), and a constant solid angle polar grid ($\Delta \mu$ constant: 32 bin from $\mu=0.01\to 1$, with another 32 from $\mu=-0.01\to -1$) and uniform grid of $\Delta \phi$ (32 bins: $0\to 2\pi$). We put $\tau=5$ between $\mu=0.01$ and the midplane so that photons cannot cross from the lower half to the upper half plane. 
The resultant density distribution is shown in Fig.\ref{fig:density}.

\subsection{Wind scattering of unpolarised input photons with different angular distributions}

The wind in the radiation-hydrodynamic simulation is set self consistently by the radiation field. In particular, the wind structure can change for different illumination patterns. However, the wind in GX 13+1 is marginally optically thick,
so this probably represents the strongest wind which can be seen as any further increase in illumination to increase the wind column density will start to shield the disc.
Hence we take this wind structure as static and explore how scattering in the wind imprints a polarisation signal on intrinsically unpolarised flux. 

We consider three different illumination patterns, one where the photons from the inner disc are isotropic, so $L(\mu)=L_0$, one where 
$L(\mu)=2\mu L_0$ (absorption dominated atmosphere), and one where $L(\mu)=\mu(1+1.8\mu)L_0/1.1$ (electron scattering: ST85) where $L_0$ is the mean (angle integrated) luminosity in all cases. 
These different polar angle flux distributions give different amounts of incident flux with the angle of the wind. 
The isotropic distribution is probably most appropriate for the bright neutron star binaries, where there is a boundary layer as well as the disc. 
The electron scattering pattern is the one expected for a disc above 1.2-1.5~keV where there is little true absorption opacity \citep{Davis2006b}, but Doppler boosting affects the illumination from a Keplarian disc, giving a pattern which is closer to the absorption dominated one. These three illumination patterns then span the range expected for all the bright low-mass X-ray binary systems. 
We use the same Monte-Carlo simulation for all three but weigh the photons differently for each illumination pattern. Hence the Poisson errors are correlated in all three cases. 

\begin{figure}
    \centering
    \includegraphics[width=0.9\hsize]{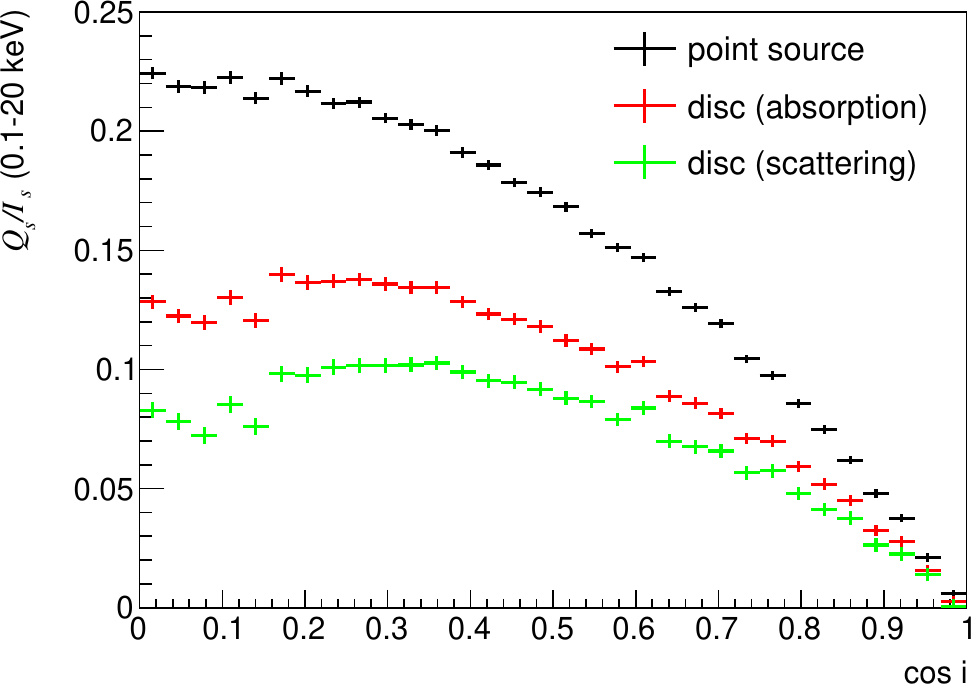}
    \caption{The polarisation degree from scattering in the wind for the three different illumination patterns. Isotropic scattering has the largest number of photons illuminating the highest inclination material, so it has a larger polarisation than the disc-like (either absorption or electron scattering) illumination patterns. Importantly, all these polarisations have a positive sign, i.e. the polarisation is perpendicular to the plane of the disc.}
    \label{fig:scattering_ps_disc}
\end{figure}

Fig.\ref{fig:scattering_ps_disc} shows
the polarisation of the scattered flux for the three illumination patterns. 
The isotropic point source gives scattered flux, which is more polarised than the disc illumination patterns, as the isotropic source has more illumination of the equatorial parts of the wind, which are at the highest inclination angles. 
The degree of polarisation decreases as the illumination becomes more polar, as it picks out higher parts of the wind which are less inclined. 
However, the most important point is that all the scattered fluxes are polarised in the same (positive) direction, i.e., polarised perpendicular to the plane of the disc.

\begin{figure}
    \centering
    \includegraphics[width=0.9\hsize]{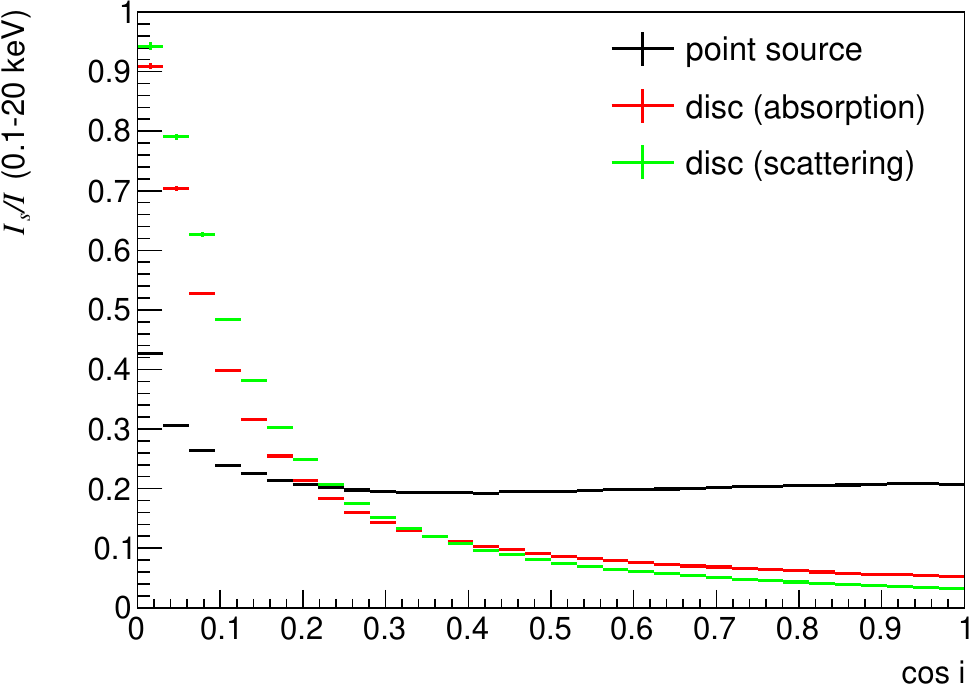}
    \caption{The ratio of the wind scattered to total (scattered plus absorbed primary) flux seen in any direction for the three illumination patterns.
    Absorption of the primary component in the wind at high inclination means that the scattered flux becomes increasingly important. 
    This is especially marked in the disc-like illumination patterns, as the primary flux is intrinsically low in the equatorial plane.}
    \label{fig:ratio}
\end{figure}

\begin{figure}
    \centering
    \includegraphics[width=0.9\hsize]{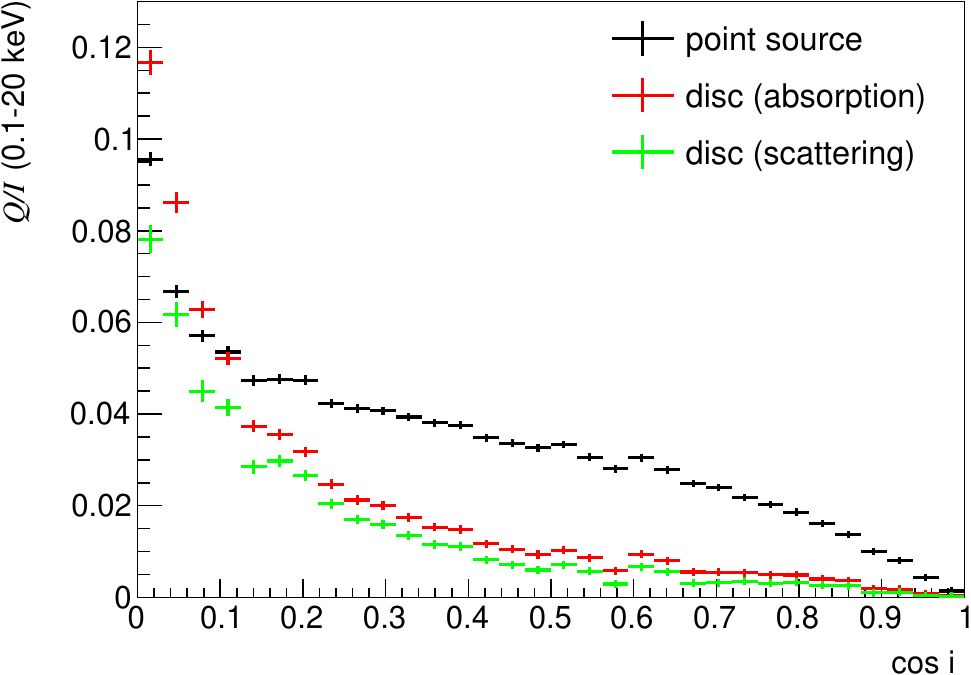}
    \caption{The polarisation degree of the total spectrum due to scattering in the wind for the three different (unpolarised) illumination patterns. This is derived from the combination of the polarisation of the scattered flux (Fig.\ref{fig:scattering_ps_disc}) diluted by the ratio of scattered to total flux (Fig.\ref{fig:ratio}). All the resultant polarisations are in the positive direction (perpendicular to the plane of the disc), as the scattered flux is the only polarised component. }
    \label{fig:total_ps_disc}
\end{figure}

\begin{figure}
    \centering
    \includegraphics[width=0.9\hsize]{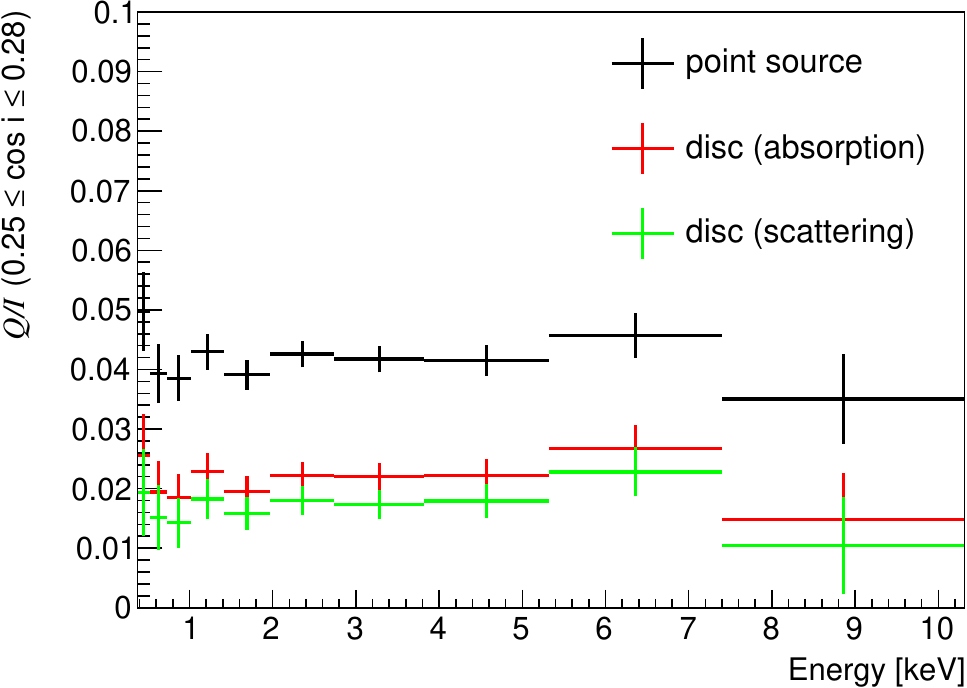}
    \caption{The polarisation degree
    of the total spectrum as a function of energy for the three different (unpolarised) illumination patterns. Each one is constant as the electron scattering cross-section is constant with energy.
    }
    \label{fig:total_ps_disc_energy}
\end{figure}

The scattered emission dominates in the subset of sources where the much brighter direct emission is obscured by the disc itself.
These are the Accretion Disc Corona sources, such as 2S0921 \citep{Tomaru2023b}.
These results predict that these sources should be strongly polarised
and with a polarisation degree, which gives a diagnostic of the angular distribution of the intrinsic source illumination pattern. 

However, in 4U1630, we see the intrinsic (primary) emission directly, so the expected polarisation from the scattered emission is diluted. 
The sum of two components, each with Stokes parameters of $I_p, Q_p$ and $I_s, Q_s$ (where $I$ is equivalent to the number of photons) gives 
$Q/I = (Q_p+Q_s)/(I_p + I_s) = (Q_p/I_p)\times I_p/I + (Q_s/I_s)\times I_s/I$.
Here we assume that the intrinsic emission is unpolarised, so $Q_p=0$.
Hence 
$Q/I = Q_s/I_s \times I_s/I$, where $I=I_p+I_s$ and $I_p = L(\mu)\exp(-\tau_\mu) $ where $\tau_\mu=1.2\sigma_T N_H(\mu)$ i.e. includes the electron scattering losses for a direct line of sight through the column. 
Fig.~\ref{fig:ratio} shows the ratio $I_s/I$ as a function of the angle for each source illumination pattern.
The scattered emission makes up over 20\% of the total emission for the isotropic source but is much less for the disc-like illumination patterns.
This is because the isotropic source has more illumination of the equatorial parts of the wind, which are the most optically thick, producing more scattered flux.

We can estimate the fraction of scattered flux analytically from $I_s=\int_{\phi=0}^{2\pi}\int_{\mu=0}^1 L(\mu) (d\Omega/2\pi) \tau_\mu$. 
These integrate to $0.3L_0$, $0.1L_0$ and $0.08L_0$, respectively, so $I_s/I_p=0.3,0.05/\mu$ and $0.088/[\mu(1+1.8\mu)]$ in the regions where absorption is negligible i.e. $\mu \geq 0.5$ (The attenuation of primary component is negligible due to the optically thin.) 
This gives a good match to the simulated ratios in Fig.~\ref{fig:ratio}, in particular predicting $I_s/(I_p+I_s)=I_s/I=0.23$ for the isotropic case, as observed at $\mu\geq 0.5$. 
The increase in optical depth for lower $\mu$ suppresses the direct component, increasing the scattered flux's contribution.
The decrease in direct component normalisation is even more marked with the disc-like illumination patterns, as these have intrinsically less flux at high inclination angles, so the scattered flux dominates more. 

We combine the ratio of scattered flux, $I_s/(I_s+I_p)$ (Fig.\ref{fig:ratio}), with the polarisation degree of the scattered photons  (Fig.\ref{fig:scattering_ps_disc}) to derive the total polarisation
in Fig.\ref{fig:total_ps_disc}. 
This is lower for the more realistic disc-like illumination patterns, both because of the lower intrinsic polarisation degree of the scattered emission (Fig.~\ref{fig:scattering_ps_disc}), and their lower contribution of the scattered to total emission (Fig.\ref{fig:ratio}). 

Fig.~\ref{fig:total_ps_disc_energy} shows the (lack of) energy dependence of this polarisation signal for $\mu=0.25$. This is expected as the simulated wind is completely ionised, so the scattering has no energy dependence.

\subsection{Wind scattering of intrinsically polarised accretion disc photons}

\begin{figure}
    \centering
    \includegraphics[width=0.9\hsize]{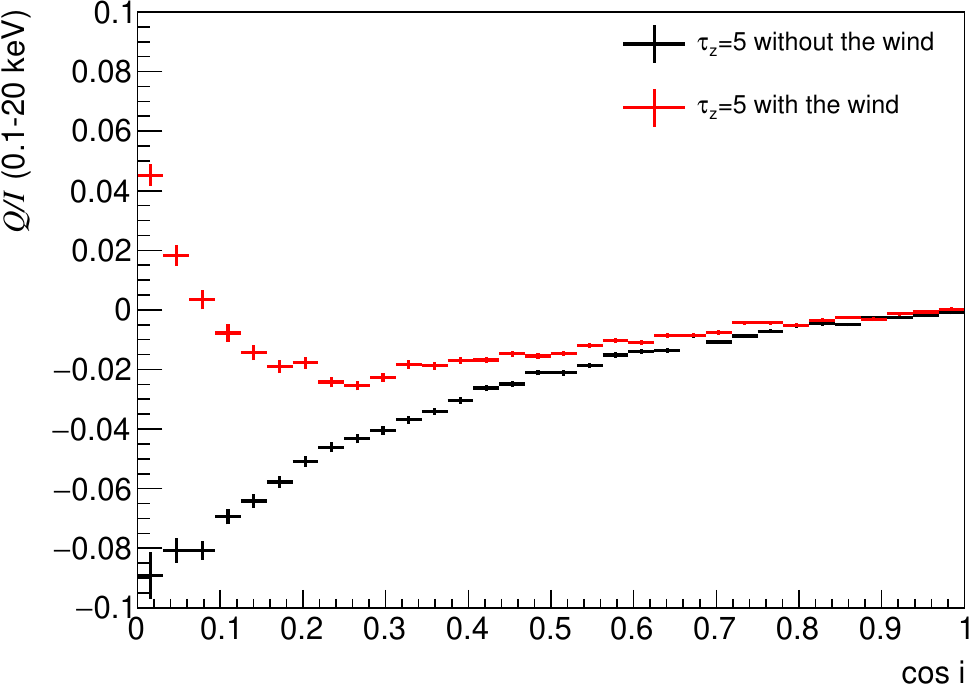}
    \caption{polarisation from the disc (black) as a function of angle, compared to polarisation from disc plus wind (red). The wind reduces the polarisation signal as the optically thin wind switches the direction of the electric vector compared to the optically thick disc (see Fig.\ref{fig:slab}) }
    \label{fig:pol_disc_wind}
\end{figure}

The disc spectrum should be intrinsically polarised from its predominantly electron scattering, plane-parallel atmosphere. 
We replace the unpolarised blackbody source used above with a thin disc structure in {\sc monaco}.
This has isotropic seed photons on the midplane with $T(R)=T_{\rm in}  (R/R_{\rm in})^{-3/4}$ 
where $T_{\rm in}=1.5$~keV and $R_{\rm in}=6 R_{\rm g} (R_{\rm g}=GM/c^2)$ to match the data. 
This is overlaid by an electron-scattering atmosphere with 
$\tau=5$ and $T_e(R)=T(R)$ to give both 
the electron scattering radiation pattern 
and polarisation degree. This is shown in 
black in Fig.~\ref{fig:pol_disc_wind}, 
and matches well to the theoretical 
Chandrasekhar result. This shows that the 
multitemperature disc gives the same 
result as the constant temperature slab 
of Fig.\ref{fig:slab}. We also confirmed that the angle dependence of this illumination is as expected for an electron-scattering atmosphere. 

The red points show the result when this 
polarised spectrum with its electron scattering angle-dependent illumination pattern scatters off the wind. 
The  polarisation of the total (scattered plus 
direct) emission is strongly reduced by 
the orthoganal  direction of the 
scattered polarisation, so the
total polarisation (wind scattering plus disc) 
at $\mu=0.25$ ($\sim 75^\circ$) decreases
from 5\% to around 3\%. This is 
much smaller than the $6-10$\% measured 
in 4U1630, showing that scattering in the wind is not a viable way to increase the polarisation signal from a standard disc.

\section{Fitting high/soft state data from 4U1630}

There is one more effect that we have not yet included, which is that the wind is not actually completely ionised as it has strong absorption lines from H- and He--like iron imprinted on it. This will suppress the direct flux at these energies, increasing the importance of the scattered flux and hence increasing the polarisation. 

We use one of the {\it NICER} datasets (ObsID: 5501010107), which was simultaneous with the {\it IXPE} observation to assess the impact of this.
We fit this in {\sc xspec}  with a model where an intrinsic simple disc spectrum ({\tt diskbb}) is absorbed by the wind.
We calculate this using {\tt pion} in {\sc spex} v3.06.01 \citep{Kaastra1996, Mao2017}. We illuminate solar abundance material by a disc spectrum (the
{\tt Dbb} model in 
{\sc spex} with temperature of 2 keV, which is  equivalent to {\tt ezdiskbb} with 1 keV in {\sc xspec}) and tabulate the results as a multiplicative model ({\tt mtable}). 
This includes the effects of both absorption by ionic lines/edges and attenuation by electron scattering. 
If the direct spectrum was the only component then this could be modelled as {\tt mtable\{~pion.fits~\}~)*diskbb }, but there is also the scattered component from the wind. We include this by using a covering fraction, $f_{\rm cov}$, so that the scattered spectrum is normalised by $1-f_{\rm cov}$. 
Hence the ratio of scattering to total flux is given by
\begin{equation}
\frac{I_s}{I_s+I_p} =\frac{1-f_{\rm cov}}{1-f_{\rm cov}+ f_{\rm cov}\exp(-\tau)}
\label{eq:ratio}
\end{equation}
All these components are affected by 
interstellar absorption, modelled with ({\tt tbabs}). The total model is {\tt tbabs*(partcov*mtable\{~pion.fits~\}~)*diskbb }.

\begin{figure}
    \centering
    \includegraphics[width=0.9\hsize,trim={1cm 2cm 1cm 8cm},clip]{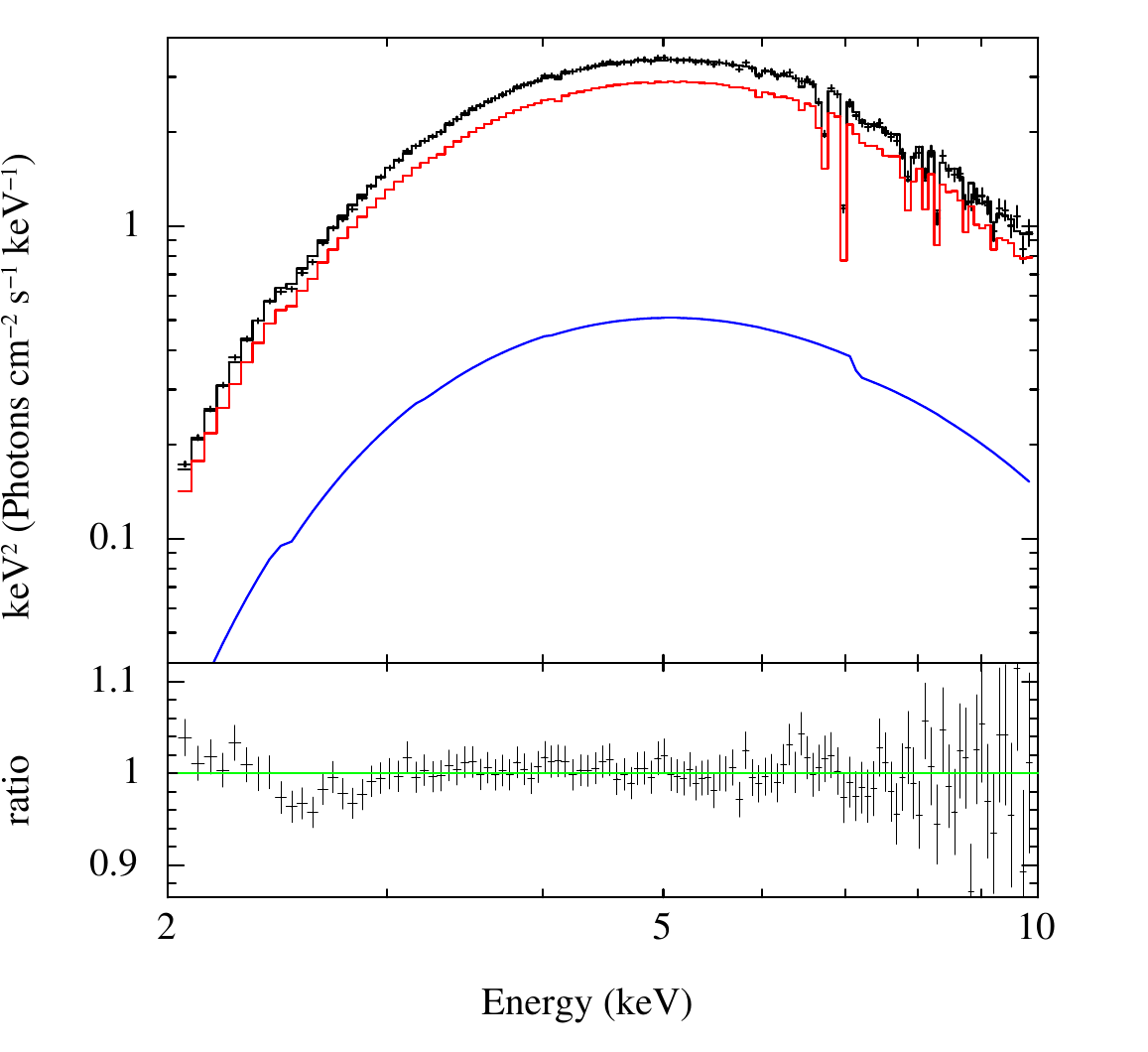}
    \caption{The upper panel shows the {\it NICER} data (black points) fit in the 2-10 keV band by a {\tt diskbb} continuum which is seen directly through the highly ionised absorber (red) and via scattering (blue). See Table~\ref{tab:xspec} for the model parameters. The lower panel shows the ratio of the model to the data, showing the fit is generally very good.
    }
    \label{fig:spec_data}
\end{figure}

\begin{figure}
    \centering
    \includegraphics[width=0.9\hsize]{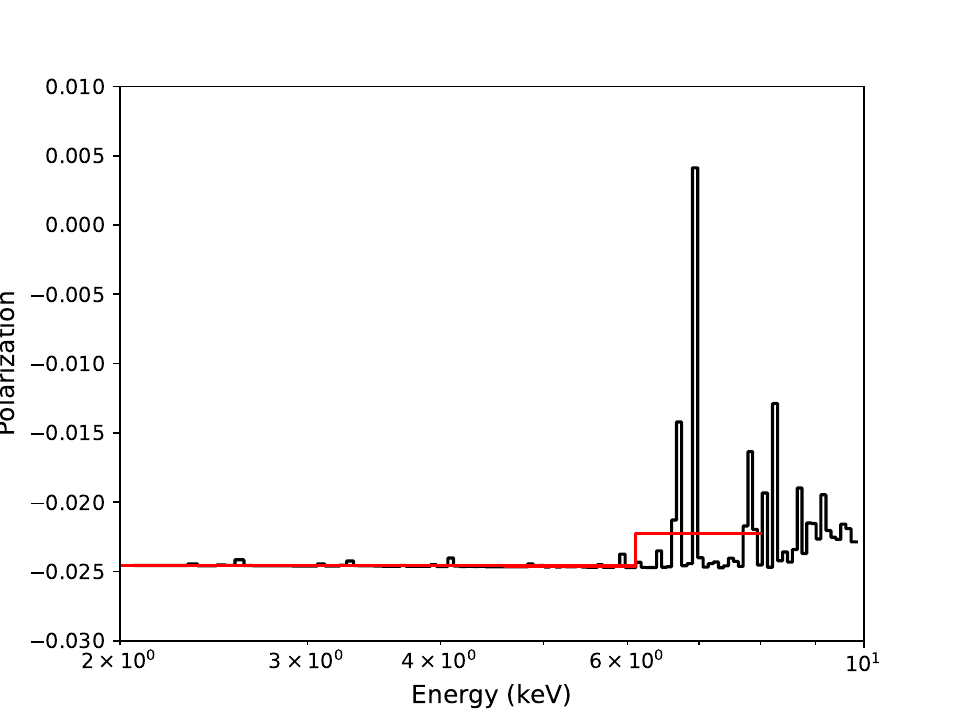}
    \caption{The black line shows the predicted polarisation degree of the total spectrum shown in Fig.~\ref{fig:spec_data} a function of energy assuming that the electron scattering dominated disc intrinsic polarisation is $-0.05$ and the scattered flux polarisation is $+0.10$. The increased importance of scattered flux in the absorption lines leads to the jump in polarisation at these energies, but when binned to the {\it IXPE} resolution (red) the effect is very small.  
    }
    \label{fig:pol_spec}
\end{figure}

We fit this to the 2-10~keV bandpass, ignoring the lower energy data as this is heavily absorbed and affected by the dust scattering in the interstellar medium along the line of sight \citep{Kalemci2018}. 
The lower energy band data does not affect the fitting of 
the ionised absorption which is of most interest here.

Fig.\ref{fig:spec_data} shows this best fit model fit to the {\it NICER} data, with the direct ($I_p$, red with absorption features) and scattered ($I_s$, blue) contributions separated out as in Eq.\ref{eq:ratio}.
The scattered component is not significantly detected (uncertainty on $f_{\rm cov}$ includes 1), but there is a strong prediction that high spectral resolution microcalorimeter data from upcoming missions such as {\it XRISM} and {\it newAthena} should see these strong absorption lines with non-zero flux at the line centre
due to the contribution of the scattered emission. 

Table \ref{tab:xspec} shows the best fit parameters. 
The disc temperature is almost identical to that seen in the {\it IXPE} observations \citep{Rawat2023}. 
The column density is almost exactly the same as predicted by our wind model, where the column along the 
line of sight at $75^\circ$ ($\mu\sim 0.25$) is $6\times 10^{22}$~cm$^{-2}$ ($N_{\rm H} = 10^{22+2.4\times 0.75}$). 
This corresponds to $\tau=0.5$. 
The fraction of scattered to total flux is 
$0.11^{+0.09}_{-0.11}$.
The best fit value indicates $\mu \sim 0.35$, but the error range 
includes our predicted ratio of $\sim 0.15-0.2$ at $\mu =0.25$ (Fig. \ref{fig:ratio}). 
Thus our wind simulation agrees with the estimated scattered flux contribution in the data.

%The corresponding angle of this is $\cos i =0.4$
%The predicted ratio is $\sim 0.15-0.2$ at $\mu 0.25$(Fig. \ref{fig:ratio}).     
%Again, this is almost exactly as predicted from our wind model in Fig.\ref{fig:ratio} for $\mu=0.25$.

\begin{table}
 \caption{Spectral fit to the 2-10~keV {\it NICER} data. 
 The {\tt pion} wind absorption has turbulent velocity fixed at 200 km/s and illuminating spectrum similar to a 1~keV {\tt diskbb}. The fit has $\chi^2/$~dof=$84/110$}
 \label{tab:xspec}
 \begin{tabular}{lcc}
  \hline
  model & parameter &  value \\
  \hline
   tbabs& $N_H/10^{22} ({\rm cm^{-2}})$ &  $7.82\pm 0.04$\\
   partcov & $f_{\rm cov}$ &   $0.93\pm 0.07$ \\
   {\tt pion} & $N_H/10^{22} ({\rm cm^{-2}})$ &  $60\pm 15$\\
    & $\log_{10} (\xi/({\rm erg~cm~s^{-1}}))$ &  $5.37\pm 0.08$\\
    &$v_{\rm out}$ (km/s)  & $840\pm 400$\\
    diskbb &$kT$ (keV) &  $1.51\pm 0.01$ \\ 
    &Norm & $230 \pm 30$\\
  \hline
 \end{tabular}
\end{table}

We fix the wind covering fraction and column to that of the simulation and estimate the additional suppression of the primary flux due to the absorption lines, and hence the increase in the fraction of scattered flux in the lines.
At $\mu=0.25$, the scattered emission has polarisation of $0.10$ while the direct flux seen from the electron scattered disc has PD of $-0.05$. 
We calculate the expected polarization as a function of energy using the {\it NICER} data resolution (black line in Fig.~\ref{fig:pol_spec}). 
This clearly shows the increase in the absorption line energies due to the increase in scattered fraction at these energies. 
However, the number of photons in the lines is quite small, so binning up to the {\it IXPE} resolution (red line) gives only a very small effect on the total polarization signal as a function of energy.
We predict that the polarization remains almost constant with energy at PD of around 0.02, which is incompatible with the observed PD, which linearly increases from 6-11\% from 2-8~keV \citep{Ratheesh2023}.

\section{Steep power law state of 4U1630}

{\it IXPE} also observed 4U1630 in a steep power law (SPL) state, where the observed luminosity is $3-4\times$ higher than that seen in the disc-dominated data \citep{Rodriguez2023, Rawat2023b}. 
There are no wind absorption features seen in the simultaneous {\it NICER} data for these data \citep{Rodriguez2023}, nor in past observations of the source in this state \citep{Hori2018,Gatuzz2019,trueba2019}. 
The increase in Compton temperature and ionisation parameter from the spectral change  is not in itself enough to make the wind completely ionised and hence invisible in absorption (\citealt{Gatuzz2019}, see also \citealt{Shidatsu2019}), but we note that the luminosity is likely super-Eddington, so radiation pressure may change the wind structure dramatically, 
perhaps making it much faster (hence less dense) and/or more equatorial (so smaller solid angle to scattering). 

If the wind column and/or opening angle has really decreased, then
the polarization signal from the wind-scattered component should also decrease. In the limit where the wind disappears, this gives a predicted drop of PD of 0.02. 
We note this is exactly the observed change in polarization between the high/soft and SPL states \citep{Rodriguez2023}, though this would also require that there is no change in intrinsic source polarization between the two very different spectra.
If the difference in polarisation is genuinely due to the presence or absence of the wind, then the intrinsic disc
polarisation must be perpendicular to the disc plane rather than parallel to the disc plane as expected from electron scattering.

\section{Wind polarization in neutron star LMXRB}

We can calculate the expected wind column density using the RHD code for any spectral shape, luminosity and disc size. The column should 
scale as $\propto L \log R_{\rm out}/0.1R_{\rm IC}$  \citep{Hori2018}
up to the point where the winds self-limit by becoming optically thick along the equatorial plane (as simulated here) and/or the luminosity goes super-Eddington (at which point the structures become highly uncertain). 

Neutron star binaries should have similar accretion discs to the black hole systems but with the addition of a bright boundary layer between the disc and solid neutron star surface. 
So far, IXPE has observed only two soft state neutron star systems with $L<L_{\rm Edd}$, namely GS 1826-238 \citep{capitanio2023} and GX 9+9 \citep{ursini2023}. 
These are very short-period (i.e. small disc) systems, so they should have very weak winds. The contribution of the wind to the polarization signal should be negligible. 

Strong winds are seen in some neutron star systems, but {\it IXPE} has not yet observed these. 
Strong blueshifted, highly ionised absorption lines are seen in the bright, high inclination, large disc system, GX 13+1. 
Our wind structure was directly calculated from this system, so our results predict the polarization signal from the wind. 
This should be energy-dependent
as the spectrum is dominated by the disc at lower energies (2-3~keV), switching to the boundary layer above this. 
The boundary layer is more isotropic, so Fig.\ref{fig:total_ps_disc_energy} predicts that this should carry a wind polarization signal of 
PD $\sim 0.04$, while the disc carries a signal with PD $\sim 0.02$.

The other large disc neutron star binary is S0921. Here the system is viewed at such a high inclination that the outer disc blocks the direct emission. 
The source is seen only via scattering in the wind, with emission lines from the highly ionised wind material dominant over absorption \citep{Tomaru2023b}. 
All the spectrum is seen via scattering, so this predicts that the polarization should increase with energy from $\sim 0.1$, where the disc illumination pattern dominates (below 3~keV), to $\sim 0.2$ where the more isotropic boundary layer dominates (see Fig.\ref{fig:scattering_ps_disc}). 
These predictions are testable with future {\it IXPE} observations, and this should enable the wind polarization to be separated from any intrinsic polarization of the disc and boundary layer. 

\section{Conclusions}

Black hole X-ray binaries in their disc-dominated state can be well fit by simple (sum of blackbodies) models of an optically thick, geometrically thin disc. 
This is the case for the high/soft state of 4U1630 observed by {\it IXPE}, but the polarisation is far larger than predicted by these models.
This motivates our exploration of the contribution to the polarisation signal from electron scattering in the accretion disc wind, which is clearly seen in this source.
However, here we show 
that the optically thin wind 
produces a scattered flux with polarisation in the opposite direction to that of the intrinsic optically thick disc emission, so the wind {\it  depolarises} the total spectrum.
Hence adding the wind to the optically thick disc cannot help explain the observed level of polarisation. 
Yet the wind is clearly present, so its scattered flux must contribute.
It is fairly easy to depolarise the optically thick disc through Faraday rotation from turbulent magnetic fields, but the wind alone predicts a polarisation signal which is too small (Fig.\ref{fig:total_ps_disc_energy}), and too constant as a function of energy. 
It seems that the only way to get the observed high level of polarisation is if the intrinsic optically thick disc can somehow be polarised in the opposite direction to the simple electron scattering expectations.
Quite how to do this remains unclear, but the answer may lie in anisotropic electrons in the photosphere \citep{Krawczynski2023}. 

\section*{Acknowledgements}

We thank Henric Krawcynski for useful discussions. RT and CD acknowledge support from the STFC consolidated grant ST/T000244/1. CD 
thanks the Lorentz centre for hosting a meeting on 'Overcoming disconnects in black hole binaries' which provided motivation to explore the polarisation properties of winds, and thanks Kavli IPMU for visiting support. 
Numerical computations were in part carried out on Cray XC50 at Center for Computational Astrophysics (CfCA), National Astronomical Observatory of Japan (NAOJ).
Numerical analyses were in part carried out on analysis servers at CfCA, NAOJ.

%%%%%%%%%%%%%%%%%%%%%%%%%%%%%%%%%%%%%%%%%%%%%%%%%%
\section*{Data Availability}

 The {\it Nicer} data are publicly available. 
 Access to the radiation transfer code is available on reasonable request from H.O.(odaka@ess.sci.osaka-u.ac.jp).

%%%%%%%%%%%%%%%%%%%% REFERENCES %%%%%%%%%%%%%%%%%%

% The best way to enter references is to use BibTeX:

\bibliographystyle{mnras}
\bibliography{library} % if your bibtex file is called example.bib

\begin{thebibliography}{}
\makeatletter
\relax
\def\mn@urlcharsother{\let\do\@makeother \do\$\do\&\do\#\do\^\do\_\do\%\do\~}
\def\mn@doi{\begingroup\mn@urlcharsother \@ifnextchar [ {\mn@doi@} {\mn@doi@[]}}
\def\mn@doi@[#1]#2{\def\@tempa{#1}\ifx\@tempa\@empty \href {http://dx.doi.org/#2} {doi:#2}\else \href {http://dx.doi.org/#2} {#1}\fi \endgroup}
\def\mn@eprint#1#2{\mn@eprint@#1:#2::\@nil}
\def\mn@eprint@arXiv#1{\href {http://arxiv.org/abs/#1} {{\tt arXiv:#1}}}
\def\mn@eprint@dblp#1{\href {http://dblp.uni-trier.de/rec/bibtex/#1.xml} {dblp:#1}}
\def\mn@eprint@#1:#2:#3:#4\@nil{\def\@tempa {#1}\def\@tempb {#2}\def\@tempc {#3}\ifx \@tempc \@empty \let \@tempc \@tempb \let \@tempb \@tempa \fi \ifx \@tempb \@empty \def\@tempb {arXiv}\fi \@ifundefined {mn@eprint@\@tempb}{\@tempb:\@tempc}{\expandafter \expandafter \csname mn@eprint@\@tempb\endcsname \expandafter{\@tempc}}}

\bibitem[\protect\citeauthoryear{{Allen}, {Schulz}, {Homan}, {Neilsen}, {Nowak}  \& {Chakrabarty}}{{Allen} et~al.}{2018}]{Allen2018}
{Allen} J.~L.,  {Schulz} N.~S.,  {Homan} J.,  {Neilsen} J.,  {Nowak} M.~A.,   {Chakrabarty} D.,  2018, \mn@doi [\apj] {10.3847/1538-4357/aac2d1}, \href {https://ui.adsabs.harvard.edu/abs/2018ApJ...861...26A} {861, 26}

\bibitem[\protect\citeauthoryear{{Angel}}{{Angel}}{1969}]{Angel1969}
{Angel} J.~R.~P.,  1969, \mn@doi [\apj] {10.1086/150185}, \href {https://ui.adsabs.harvard.edu/abs/1969ApJ...158..219A} {158, 219}

\bibitem[\protect\citeauthoryear{Begelman, McKee  \& Shields}{Begelman et~al.}{1983}]{Begelman1983a}
Begelman M.~C.,  McKee C.~F.,   Shields G.~A.,  1983, \mn@doi [\apj] {10.1086/161178}, 271, 70

\bibitem[\protect\citeauthoryear{{Capitanio} et~al.,}{{Capitanio} et~al.}{2023}]{capitanio2023}
{Capitanio} F.,  et~al., 2023, \mn@doi [\apj] {10.3847/1538-4357/acae88}, \href {https://ui.adsabs.harvard.edu/abs/2023ApJ...943..129C} {943, 129}

\bibitem[\protect\citeauthoryear{{Chandrasekhar}}{{Chandrasekhar}}{1960}]{Chandrasekhar1960}
{Chandrasekhar} S.,  1960, {Radiative transfer}

\bibitem[\protect\citeauthoryear{{Connors}, {Piran}  \& {Stark}}{{Connors} et~al.}{1980}]{Connors1980}
{Connors} P.~A.,  {Piran} T.,   {Stark} R.~F.,  1980, \mn@doi [\apj] {10.1086/157627}, \href {https://ui.adsabs.harvard.edu/abs/1980ApJ...235..224C} {235, 224}

\bibitem[\protect\citeauthoryear{{Davis}, {Done}  \& {Blaes}}{{Davis} et~al.}{2006}]{Davis2006b}
{Davis} S.~W.,  {Done} C.,   {Blaes} O.~M.,  2006, \mn@doi [\apj] {10.1086/505386}, \href {https://ui.adsabs.harvard.edu/\#abs/2006ApJ...647..525D} {647, 525}

\bibitem[\protect\citeauthoryear{{D{\'{i}}az Trigo}, Migliari  \& Guainazzi}{{D{\'{i}}az Trigo} et~al.}{2014}]{DiazTrigo2014}
{D{\'{i}}az Trigo} M.,  Migliari S.,   Guainazzi M.,  2014, \mn@doi [\aap] {10.1051/0004-6361/201424554}, 76, 1

\bibitem[\protect\citeauthoryear{{Done} \& {Gierli{\'n}ski}}{{Done} \& {Gierli{\'n}ski}}{2005}]{Done2005}
{Done} C.,  {Gierli{\'n}ski} M.,  2005, \mn@doi [\mnras] {10.1111/j.1365-2966.2005.09555.x}, \href {https://ui.adsabs.harvard.edu/abs/2005MNRAS.364..208D} {364, 208}

\bibitem[\protect\citeauthoryear{{Done}, {Tomaru}  \& {Takahashi}}{{Done} et~al.}{2018}]{Done2018}
{Done} C.,  {Tomaru} R.,   {Takahashi} T.,  2018, \mn@doi [\mnras] {10.1093/mnras/stx2400}, \href {http://adsabs.harvard.edu/abs/2018MNRAS.473..838D} {473, 838}

\bibitem[\protect\citeauthoryear{{Dovciak}, {Steiner}, {Krawczynski}  \& {Svoboda}}{{Dovciak} et~al.}{2023}]{Dovciak2023}
{Dovciak} M.,  {Steiner} J.~F.,  {Krawczynski} H.,   {Svoboda} J.,  2023, The Astronomer's Telegram, \href {https://ui.adsabs.harvard.edu/abs/2023ATel16084....1D} {16084, 1}

\bibitem[\protect\citeauthoryear{{Dov{\v{c}}iak}, {Muleri}, {Goosmann}, {Karas}  \& {Matt}}{{Dov{\v{c}}iak} et~al.}{2008}]{Dovciak2008}
{Dov{\v{c}}iak} M.,  {Muleri} F.,  {Goosmann} R.~W.,  {Karas} V.,   {Matt} G.,  2008, \mn@doi [\mnras] {10.1111/j.1365-2966.2008.13872.x}, \href {https://ui.adsabs.harvard.edu/abs/2008MNRAS.391...32D} {391, 32}

\bibitem[\protect\citeauthoryear{{Gatuzz}, {D{\'\i}az Trigo}, {Miller-Jones}  \& {Migliari}}{{Gatuzz} et~al.}{2019}]{Gatuzz2019}
{Gatuzz} E.,  {D{\'\i}az Trigo} M.,  {Miller-Jones} J.~C.~A.,   {Migliari} S.,  2019, \mn@doi [\mnras] {10.1093/mnras/sty2850}, \href {https://ui.adsabs.harvard.edu/abs/2019MNRAS.482.2597G} {482, 2597}

\bibitem[\protect\citeauthoryear{{Gianolli} et~al.,}{{Gianolli} et~al.}{2023}]{Gianolli2023}
{Gianolli} V.~E.,  et~al., 2023, \mn@doi [\mnras] {10.1093/mnras/stad1697}, \href {https://ui.adsabs.harvard.edu/abs/2023MNRAS.523.4468G} {523, 4468}

\bibitem[\protect\citeauthoryear{{Hori}, {Ueda}, {Done}, {Shidatsu}  \& {Kubota}}{{Hori} et~al.}{2018}]{Hori2018}
{Hori} T.,  {Ueda} Y.,  {Done} C.,  {Shidatsu} M.,   {Kubota} A.,  2018, \mn@doi [\apj] {10.3847/1538-4357/aaea5e}, \href {https://ui.adsabs.harvard.edu/abs/2018ApJ...869..183H} {869, 183}

\bibitem[\protect\citeauthoryear{{Kaastra}, {Mewe}  \& {Nieuwenhuijzen}}{{Kaastra} et~al.}{1996}]{Kaastra1996}
{Kaastra} J.~S.,  {Mewe} R.,   {Nieuwenhuijzen} H.,  1996, in UV and X-ray Spectroscopy of Astrophysical and Laboratory Plasmas. pp 411--414

\bibitem[\protect\citeauthoryear{{Kalemci}, {Maccarone}  \& {Tomsick}}{{Kalemci} et~al.}{2018}]{Kalemci2018}
{Kalemci} E.,  {Maccarone} T.~J.,   {Tomsick} J.~A.,  2018, \mn@doi [\apj] {10.3847/1538-4357/aabcd3}, \href {https://ui.adsabs.harvard.edu/abs/2018ApJ...859...88K} {859, 88}

\bibitem[\protect\citeauthoryear{{Krawczynski} et~al.,}{{Krawczynski} et~al.}{2022}]{Krawczynski2022}
{Krawczynski} H.,  et~al., 2022, \mn@doi [Science] {10.1126/science.add5399}, \href {https://ui.adsabs.harvard.edu/abs/2022Sci...378..650K} {378, 650}

\bibitem[\protect\citeauthoryear{{Krawczynski}, {Yuan}, {Chen}, {Rodriguez Cavero}, {Hu}, {Gau}, {Steiner}  \& {Dov{\v{c}}iak}}{{Krawczynski} et~al.}{2023}]{Krawczynski2023}
{Krawczynski} H.,  {Yuan} Y.,  {Chen} A.~Y.,  {Rodriguez Cavero} N.,  {Hu} K.,  {Gau} E.,  {Steiner} J.~F.,   {Dov{\v{c}}iak} M.,  2023, \mn@doi [arXiv e-prints] {10.48550/arXiv.2307.13141}, \href {https://ui.adsabs.harvard.edu/abs/2023arXiv230713141K} {p. arXiv:2307.13141}

\bibitem[\protect\citeauthoryear{Kubota et~al.,}{Kubota et~al.}{2007}]{Kubota2007}
Kubota A.,  et~al., 2007, \mn@doi [\pasj] {10.1093/pasj/59.sp1.S185}, 59, S185

\bibitem[\protect\citeauthoryear{{Kubota}, {Odaka}, {Tamagawa}  \& {Nakano}}{{Kubota} et~al.}{2018}]{Kubota2018}
{Kubota} M.,  {Odaka} H.,  {Tamagawa} T.,   {Nakano} T.,  2018, \mn@doi [\apjl] {10.3847/2041-8213/aaef76}, \href {https://ui.adsabs.harvard.edu/abs/2018ApJ...868L..26K} {868, L26}

\bibitem[\protect\citeauthoryear{{Kushwaha}, {Jayasurya}, {Agrawal}  \& {Nandi}}{{Kushwaha} et~al.}{2023}]{Kushwaha2023}
{Kushwaha} A.,  {Jayasurya} K.~M.,  {Agrawal} V.~K.,   {Nandi} A.,  2023, \mn@doi [\mnras] {10.1093/mnrasl/slad070}, \href {https://ui.adsabs.harvard.edu/abs/2023MNRAS.524L..15K} {524, L15}

\bibitem[\protect\citeauthoryear{{Kuulkers}, {Wijnands}, {Belloni}, {M{\'e}ndez}, {van der Klis}  \& {van Paradijs}}{{Kuulkers} et~al.}{1998}]{Kuulkers1998}
{Kuulkers} E.,  {Wijnands} R.,  {Belloni} T.,  {M{\'e}ndez} M.,  {van der Klis} M.,   {van Paradijs} J.,  1998, \mn@doi [\apj] {10.1086/305248}, \href {https://ui.adsabs.harvard.edu/abs/1998ApJ...494..753K} {494, 753}

\bibitem[\protect\citeauthoryear{{Mahmoud} \& {Done}}{{Mahmoud} \& {Done}}{2020}]{Mahmoud2020}
{Mahmoud} R.~D.,  {Done} C.,  2020, \mn@doi [\mnras] {10.1093/mnras/stz3196}, \href {https://ui.adsabs.harvard.edu/abs/2020MNRAS.491.5126M} {491, 5126}

\bibitem[\protect\citeauthoryear{{Mao}, {Kaastra}, {Mehdipour}, {Raassen}, {Gu}  \& {Miller}}{{Mao} et~al.}{2017}]{Mao2017}
{Mao} J.,  {Kaastra} J.~S.,  {Mehdipour} M.,  {Raassen} A.~J.~J.,  {Gu} L.,   {Miller} J.~M.,  2017, \mn@doi [\aap] {10.1051/0004-6361/201731378}, \href {https://ui.adsabs.harvard.edu/abs/2017A&A...607A.100M} {607, A100}

\bibitem[\protect\citeauthoryear{{Matt}, {Feroci}, {Rapisarda}  \& {Costa}}{{Matt} et~al.}{1996}]{Matt1996}
{Matt} G.,  {Feroci} M.,  {Rapisarda} M.,   {Costa} E.,  1996, \mn@doi [Radiation Physics and Chemistry] {10.1016/0969-806X(95)00472-A}, \href {https://ui.adsabs.harvard.edu/abs/1996RaPC...48..403M} {48, 403}

\bibitem[\protect\citeauthoryear{{Mikusincova}, {Dovciak}, {Bursa}, {Lalla}, {Matt}, {Svoboda}, {Taverna}  \& {Zhang}}{{Mikusincova} et~al.}{2023}]{Mikusincova2023}
{Mikusincova} R.,  {Dovciak} M.,  {Bursa} M.,  {Lalla} N.~D.,  {Matt} G.,  {Svoboda} J.,  {Taverna} R.,   {Zhang} W.,  2023, \mn@doi [\mnras] {10.1093/mnras/stad077}, \href {https://ui.adsabs.harvard.edu/abs/2023MNRAS.519.6138M} {519, 6138}

\bibitem[\protect\citeauthoryear{{Neilsen}, {Coriat}, {Fender}, {Lee}, {Ponti}, {Tzioumis}, {Edwards}  \& {Broderick}}{{Neilsen} et~al.}{2014}]{Neilsen2014}
{Neilsen} J.,  {Coriat} M.,  {Fender} R.,  {Lee} J.~C.,  {Ponti} G.,  {Tzioumis} A.~K.,  {Edwards} P.~G.,   {Broderick} J.~W.,  2014, \mn@doi [\apjl] {10.1088/2041-8205/784/1/L5}, \href {https://ui.adsabs.harvard.edu/abs/2014ApJ...784L...5N} {784, L5}

\bibitem[\protect\citeauthoryear{Odaka, Aharonian, Watanabe, Tanaka, Khangulyan  \& Takahashi}{Odaka et~al.}{2011}]{Odaka2011}
Odaka H.,  Aharonian F.,  Watanabe S.,  Tanaka Y.,  Khangulyan D.,   Takahashi T.,  2011, \mn@doi [\apj] {10.1088/0004-637X/740/2/103}, 740, 103

\bibitem[\protect\citeauthoryear{{Odaka}, {Khangulyan}, {Tanaka}, {Watanabe}, {Takahashi}  \& {Makishima}}{{Odaka} et~al.}{2014}]{Odaka2014}
{Odaka} H.,  {Khangulyan} D.,  {Tanaka} Y.~T.,  {Watanabe} S.,  {Takahashi} T.,   {Makishima} K.,  2014, \mn@doi [\apj] {10.1088/0004-637X/780/1/38}, \href {https://ui.adsabs.harvard.edu/abs/2014ApJ...780...38O} {780, 38}

\bibitem[\protect\citeauthoryear{Ponti, Fender, Begelman, Dunn, Neilsen  \& Coriat}{Ponti et~al.}{2012}]{ponti2012}
Ponti G.,  Fender R.~P.,  Begelman M.~C.,  Dunn R. J.~H.,  Neilsen J.,   Coriat M.,  2012, \mn@doi [\mnras: Letters] {10.1111/j.1745-3933.2012.01224.x}, 422, 11

\bibitem[\protect\citeauthoryear{{Ratheesh} et~al.,}{{Ratheesh} et~al.}{2023}]{Ratheesh2023}
{Ratheesh} A.,  et~al., 2023, \mn@doi [arXiv e-prints] {10.48550/arXiv.2304.12752}, \href {https://ui.adsabs.harvard.edu/abs/2023arXiv230412752R} {p. arXiv:2304.12752}

\bibitem[\protect\citeauthoryear{{Rawat}, {Garg}  \& {M{\'e}ndez}}{{Rawat} et~al.}{2023a}]{Rawat2023b}
{Rawat} D.,  {Garg} A.,   {M{\'e}ndez} M.,  2023a, \mn@doi [\mnras] {10.1093/mnras/stad2327}, \href {https://ui.adsabs.harvard.edu/abs/2023MNRAS.525..661R} {525, 661}

\bibitem[\protect\citeauthoryear{{Rawat}, {Garg}  \& {M{\'e}ndez}}{{Rawat} et~al.}{2023b}]{Rawat2023}
{Rawat} D.,  {Garg} A.,   {M{\'e}ndez} M.,  2023b, \mn@doi [\apjl] {10.3847/2041-8213/acd77b}, \href {https://ui.adsabs.harvard.edu/abs/2023ApJ...949L..43R} {949, L43}

\bibitem[\protect\citeauthoryear{{Rodriguez Cavero} et~al.,}{{Rodriguez Cavero} et~al.}{2023}]{Rodriguez2023}
{Rodriguez Cavero} N.,  et~al., 2023, \mn@doi [arXiv e-prints] {10.48550/arXiv.2305.10630}, \href {https://ui.adsabs.harvard.edu/abs/2023arXiv230510630R} {p. arXiv:2305.10630}

\bibitem[\protect\citeauthoryear{{Shidatsu} \& {Done}}{{Shidatsu} \& {Done}}{2019}]{Shidatsu2019}
{Shidatsu} M.,  {Done} C.,  2019, arXiv e-prints, \href {https://ui.adsabs.harvard.edu/abs/2019arXiv190602469S} {p. arXiv:1906.02469}

\bibitem[\protect\citeauthoryear{{Sugimoto}, {Mihara}, {Kitamoto}, {Matsuoka}, {Sugizaki}, {Negoro}, {Nakahira}  \& {Makishima}}{{Sugimoto} et~al.}{2016}]{Sugimoto2016}
{Sugimoto} J.,  {Mihara} T.,  {Kitamoto} S.,  {Matsuoka} M.,  {Sugizaki} M.,  {Negoro} H.,  {Nakahira} S.,   {Makishima} K.,  2016, \mn@doi [\pasj] {10.1093/pasj/psw004}, \href {https://ui.adsabs.harvard.edu/abs/2016PASJ...68S..17S} {68, S17}

\bibitem[\protect\citeauthoryear{{Sunyaev} \& {Titarchuk}}{{Sunyaev} \& {Titarchuk}}{1985}]{Sunyaev1985}
{Sunyaev} R.~A.,  {Titarchuk} L.~G.,  1985, \aap, \href {https://ui.adsabs.harvard.edu/abs/1985A&A...143..374S} {143, 374}

\bibitem[\protect\citeauthoryear{{Tomaru}, {Done}, {Ohsuga}, {Nomura}  \& {Takahashi}}{{Tomaru} et~al.}{2019}]{Tomaru2019}
{Tomaru} R.,  {Done} C.,  {Ohsuga} K.,  {Nomura} M.,   {Takahashi} T.,  2019, \mn@doi [\mnras] {10.1093/mnras/stz2738}, \href {https://ui.adsabs.harvard.edu/abs/2019MNRAS.490.3098T} {490, 3098}

\bibitem[\protect\citeauthoryear{{Tomaru}, {Done}, {Ohsuga}, {Odaka}  \& {Takahashi}}{{Tomaru} et~al.}{2020a}]{Tomaru2020}
{Tomaru} R.,  {Done} C.,  {Ohsuga} K.,  {Odaka} H.,   {Takahashi} T.,  2020a, \mn@doi [\mnras] {10.1093/mnras/staa961}, \href {https://ui.adsabs.harvard.edu/abs/2020MNRAS.494.3413T} {494, 3413}

\bibitem[\protect\citeauthoryear{{Tomaru}, {Done}, {Ohsuga}, {Odaka}  \& {Takahashi}}{{Tomaru} et~al.}{2020b}]{Tomaru2020b}
{Tomaru} R.,  {Done} C.,  {Ohsuga} K.,  {Odaka} H.,   {Takahashi} T.,  2020b, \mn@doi [\mnras] {10.1093/mnras/staa2254}, \href {https://ui.adsabs.harvard.edu/abs/2020MNRAS.497.4970T} {497, 4970}

\bibitem[\protect\citeauthoryear{{Tomaru}, {Done}  \& {Mao}}{{Tomaru} et~al.}{2023a}]{Tomaru2023a}
{Tomaru} R.,  {Done} C.,   {Mao} J.,  2023a, \mn@doi [\mnras] {10.1093/mnras/stac3210}, \href {https://ui.adsabs.harvard.edu/abs/2023MNRAS.518.1789T} {518, 1789}

\bibitem[\protect\citeauthoryear{{Tomaru}, {Chris}, {Odaka}  \& {Tanimoto}}{{Tomaru} et~al.}{2023b}]{Tomaru2023b}
{Tomaru} R.,  {Chris} D.,  {Odaka} H.,   {Tanimoto} A.,  2023b, \mn@doi [\mnras] {10.1093/mnras/stad1637}, \href {https://ui.adsabs.harvard.edu/abs/2023MNRAS.523.3441T} {523, 3441}

\bibitem[\protect\citeauthoryear{{Tomsick}, {Lapshov}  \& {Kaaret}}{{Tomsick} et~al.}{1998}]{Tomsick1998}
{Tomsick} J.~A.,  {Lapshov} I.,   {Kaaret} P.,  1998, \mn@doi [\apj] {10.1086/305240}, \href {https://ui.adsabs.harvard.edu/abs/1998ApJ...494..747T} {494, 747}

\bibitem[\protect\citeauthoryear{{Trueba}, {Miller}, {Kaastra}, {Zoghbi}, {Fabian}, {Kallman}, {Proga}  \& {Raymond}}{{Trueba} et~al.}{2019}]{trueba2019}
{Trueba} N.,  {Miller} J.~M.,  {Kaastra} J.,  {Zoghbi} A.,  {Fabian} A.~C.,  {Kallman} T.,  {Proga} D.,   {Raymond} J.,  2019, \mn@doi [\apj] {10.3847/1538-4357/ab4f70}, \href {https://ui.adsabs.harvard.edu/abs/2019ApJ...886..104T} {886, 104}

\bibitem[\protect\citeauthoryear{{Ursini} et~al.,}{{Ursini} et~al.}{2023}]{ursini2023}
{Ursini} F.,  et~al., 2023, \mn@doi [\aap] {10.1051/0004-6361/202346541}, \href {https://ui.adsabs.harvard.edu/abs/2023A&A...676A..20U} {676, A20}

\bibitem[\protect\citeauthoryear{{Veledina} et~al.,}{{Veledina} et~al.}{2023}]{Veledina2023}
{Veledina} A.,  et~al., 2023, \mn@doi [arXiv e-prints] {10.48550/arXiv.2303.01174}, \href {https://ui.adsabs.harvard.edu/abs/2023arXiv230301174V} {p. arXiv:2303.01174}

\bibitem[\protect\citeauthoryear{{Weisskopf} et~al.,}{{Weisskopf} et~al.}{2016}]{Weisskopf2016}
{Weisskopf} M.~C.,  et~al., 2016, in {den Herder} J.-W.~A.,  {Takahashi} T.,   {Bautz} M.,  eds,  Society of Photo-Optical Instrumentation Engineers (SPIE) Conference Series Vol. 9905, Space Telescopes and Instrumentation 2016: Ultraviolet to Gamma Ray. p. 990517, \mn@doi{10.1117/12.2235240}

\bibitem[\protect\citeauthoryear{{Zdziarski}, {Veledina}, {Szanecki}, {Green}, {Bright}  \& {Williams}}{{Zdziarski} et~al.}{2023}]{Zdziarski2023}
{Zdziarski} A.~A.,  {Veledina} A.,  {Szanecki} M.,  {Green} D.~A.,  {Bright} J.~S.,   {Williams} D. R.~A.,  2023, \mn@doi [\apjl] {10.3847/2041-8213/ace2c9}, \href {https://ui.adsabs.harvard.edu/abs/2023ApJ...951L..45Z} {951, L45}

\makeatother
\end{thebibliography}

% Alternatively you could enter them by hand, like this:
% This method is tedious and prone to error if you have lots of references
%\begin{thebibliography}{99}
%\bibitem[\protect\citeauthoryear{Author}{2012}]{Author2012}
%Author A.~N., 2013, Journal of Improbable Astronomy, 1, 1
%\bibitem[\protect\citeauthoryear{Others}{2013}]{Others2013}
%Others S., 2012, Journal of Interesting Stuff, 17, 198
%\end{thebibliography}

%%%%%%%%%%%%%%%%%%%%%%%%%%%%%%%%%%%%%%%%%%%%%%%%%%

%%%%%%%%%%%%%%%%% APPENDICES %%%%%%%%%%%%%%%%%%%%%

%%%%%%%%%%%%%%%%%%%%%%%%%%%%%%%%%%%%%%%%%%%%%%%%%%

% Don't change these lines
\bsp	% typesetting comment
\label{lastpage}
\end{document}